\documentclass[11pt]{article}
\usepackage{graphicx}
\usepackage{amsmath}
\usepackage[usenames,dvipsnames]{color}
\usepackage{setspace}
\usepackage{enumitem}
\usepackage{lipsum}

\usepackage{amssymb}
\usepackage{amsmath}
\usepackage{mathrsfs}
\usepackage{rotating}

\def\st{\mbox{s.~t.}}

\title{An Integer Programming Formulation of the
Key Management Problem in Wireless Sensor Networks}

\author{
Guanglin Xu\thanks{Institute for Mathematics and its Applications, University of Minnesota,
Minneapolis, MN, 55455, USA. Email: {\tt gxu@umn.edu}.}%
\ \ \ \ \ \ \
Alexander Semenov\thanks{Information Technology, University of Jyv{\"a}skyl{\"a}, Jyv{\"a}skyl{\"a}, Finland. 
Email: {\tt alexander.v.semenov@jyu.fi}.}%
\ \ \ \ \ \ \
Maciej Rysz\thanks{Department of Industrial and Systems Engineering, University of Florida,
              Gainesville, FL, USA. 
Email: {\tt mwrysz@ufl.edu}.}%
}

\date{\today}

\begin{document}
\maketitle

\begin{abstract}

\noindent With the advent of modern communications systems, 
much attention has been put on developing methods for securely 
transferring information between constituents of  wireless sensor networks. 
To this effect, we introduce a mathematical programming formulation for the 
key management problem, which broadly serves as a mechanism for  
encrypting communications. In particular, an integer programming model 
of the $q$-Composite scheme is proposed and utilized to distribute keys 
among nodes of a network whose topology is known. Numerical experiments 
demonstrating the effectiveness of the proposed model are conducted using 
using a well-known optimization solver package. An illustrative example 
depicting an optimal encryption for a small-scale network is also presented.

\mbox{}

\end{abstract}

\begin{singlespace}

\section{Introduction}
\label{S:1}

A desirable feature to incorporate in a wireless sensor network (WSN) is the capability to securely transfer information among its constituents. Achieving this requires the advancement of complex encryption strategies that can guarantee secure communications when one or more network sensors are compromised (e.g., hacked). Although designing communication schemes under this scope does not necessarily require that the underlying  network be impervious to any form of attack, it should induce a high tolerance against the interception of system-wide communications whenever a given number of sensors are infiltrated.

WSNs are composed of communicating sensors (nodes) embedded in electronics and software that are capable of collecting, 
exchanging and/or disseminating information. They have found a 
wide-range of applications in many domains and can vastly range 
in size and connectivity characteristics. Among others, they have been utilized for monitoring structural ``health'' of bridges and other infrastructure, sensing seismic events, traffic control,  diagnosing health of the elderly, pipeline monitoring, 
agricultural/terrestrial applications, oceanography,  military, and so on 
(e.g., see~\cite{Fundamentals.2010}, \cite{Kim_2007}, \cite{he_achieving_2006}). 
For instance, \cite{Kim_2007} considered a sensor network with fifty-nine nodes that was deployed on the Golden Gate Bridge, with the primary task of collecting ambient vibrations of the whole structure.
In another article, \cite{werner_allen_1607983} studied the application of a sixteen-node WSN to monitor volcanic activity, thus enabling scientific studies that wouldn't be feasible with conventional instrumentation. The sensors nodes that were used had a program memory of 48 KB, 10 KB of static RAM, and 1 MB of external flash memory. Indeed, \cite{Eschenauer_2002} states that the memory size for wireless sensor nodes generally range from 8Kb to 80 Kb of program memory, and 512 bytes to 5.12 KB of data memory.

Due to the fact that WSNs are often deployed in volatile operational environments in which nodes may be physically or virtually captured, it is important to develop communication encryption strategies that are resilient in such settings.
Establishing secure communications typically involves distributing secret cryptographic \textit{keys} that are used to encrypt and decrypt information transfered between network nodes that are within \textit{proximity} of one another. Particularly, a given pair of nodes within proximity must share a certain number of keys in order to directly communicate in a secure manner. 
Proximities between sensors may, for example, be determined by various factors like physical distances, geographical obstacles, signal interference, and so on. 
Then, the problem of distributing keys to sensors, also known as the \textit{key management problem} (KMP) in the cyber security literature, comprises strategies for assigning secret cryptographic keys to nodes so as to enable all pairs of nodes to directly or indirectly communicate.  
Additionally, in WSNs the key distribution process is constrained by  
nodes' \textit{capacities} and computation power (if any), thereby limiting 
the number of keys that they can store and process in a reasonable amount of time, respectively.
Overall, well-constructed key distribution schemes can be applied to make communications between the sensors secure from interception and modification.


Many key distribution protocols assume that the topology of the WSN is not known 
in advance and apply a key pre-distribution scheme before the sensors are deployed. 
One of the most widely utilized pre-distribution techniques relies on synchronous 
encryption. Namely, keys are drawn from a \textit{key pool} and assigned to the 
nodes. Any node's memory capacity, in turn, limits the number of keys it can store 
to a small fraction of total keys in the key pool. Then, in order for a pair of nodes to 
communicate, they must share a prescribed number of keys according to a defined 
scheme. Surveys of key distribution schemes may be found in~\cite{chen_survey_2014}, 
\cite{ruj_pairwise_2013}, and~\cite{Simplicio_2010}. Several existing approaches include:

\begin{itemize}
  \item \textbf{Single key scheme}. Each node is preloaded with the same secret key 
and all the sensors use it for encryption of their communication. Although this scheme
is resource efficient in that only one key is stored in each sensor's memory, it fails to 
deliver a resilient security profile due to the fact that if one node is compromised, then 
the all network communications can be intercepted.

\item \textbf{Naive fully pairwise scheme}. Each node is preloaded with one unique key 
for every other sensor in the WSN, thereby each pair of nodes can share a unique key. 
This method offers a high level of security, as any compromised node only reveals 
communications with sensors within its ``proximity''. However, it is resource inefficient 
given that the memory capacity limits may inhibit sensors from storing a large number 
of keys, particularly in cases when the WSNs are large. 

\item \textbf{An Erd\"os-R\'enyi graph-based method~\cite{Eschenauer_2002}}.
Sensors are preloaded with a \textit{key ring} comprising of a subset of keys that are randomly drawn without replacement from a 
randomly generated key pool. Once the sensors are randomly deployed (e.g., dropped from an airplane) and 
the WSN structure is known, two nodes with shared keys may establish a link. It was 
shown that inter-network communications (i.e., connectivity) can be guaranteed with a 
high probability for certain network configurations and predefined key ring and 
key pools sizes.

\item \textbf{$q$-Composite scheme}. Similar to the described method~\cite{Eschenauer_2002}, 
yet two nodes must share $q$ common keys in order to establish a communication link.

\item \textbf{Multiple space scheme}. Combination of~\cite{Blom1985} and~\cite{Eschenauer_2002}, 
where randomly scattered nodes establish communication channels if they have a common key space (see \cite{Blom1985}), 

\item \textbf{Polynomial pool-based scheme} \cite{Liu_2003}. This approach combines the techniques in \cite{Eschenauer_2002} 
and \cite{Blundo1993}. Each node is scattered randomly according to \cite{Eschenauer_2002}. However, instead of storing crytographic keys, sets of bivariate symmetric polynomials are stored, which, in turn, are used in a key exchange scheme defined in  \cite{Blundo1993}.

\item \textbf{Combinatorial scheme} \cite{Camtepe_2007}. A deterministic approach for key distribution based on combinatorial design theory.

\end{itemize}

In this work, we propose an optimization-based modeling 
framework for finding optimal key management policies that produce a desired level 
of communication security.
The proposed model offers encryption strategies that provide a user-defined level of security while considering the underlying WSN's topology 
and the limited memory capacities of nodes relative to the size of the key pool. 
As a demonstrative example, it employs security requirements that eliminate the overuse of any particular key by limiting the  number of times it can be assigned among all the sensor nodes, and  the number of times it can be assigned within the neighborhood of any particular node. This prevents overly repetitive establishment of communications via the same key, which is a highly desirable security feature.  

As an initial application, emphasis is put on the $q$-composite key pre-distribution scheme where two nodes within proximity of one another -- defined by an edge connecting them -- must share at least $q$ common keys in order to communicate. We assume that either the topology of the sensor network is known prior to deployment of the sensors; or that the topology is known after deployment, in which case, the KMP can be solved to redistribute the keys in the network. Such networks typically form in situations when the sensors are randomly deployed or scattered, and subsequently need to establish secure communications among nodes within proximities. 

\subsection{Our contributions}
This study introduces a novel mathematical programming application for identifying optimal key management strategies in situations when the topology of the underlying WSN is known.  Particularly, a quadratically constrained integer programming formulation of the {\em $q$-Composition key distribution} scheme is introduced. An equivalent integer linear reformulation is solved using a well-known off-the-shelf solver. Numerical experiments are conducted on randomly generated graphs of various configurations.  To the authors' knowledge, this is the first study to utilize integer programming techniques to address the key management problem.

\subsection{Organization of the paper}

The remaining sections are organized as follows. In Section \ref{sec:problem}, we
discuss the $q$-{\em Composite key distribution scheme} and 
propose a technique for optimally assigning keys to the network nodes.
The corresponding key management problem is then formulated as an integer program with quadratic constraints, and subsequently reformulated as an integer linear 
program. 
In Section \ref{sec:exp}, numerical experiments demonstrating the effectiveness of the proposed approach on networks with various topological configurations for cases when $q=1$ and $q=2$ are conducted.
Finally, concluding remarks about future extensions for optimization-based key management problems are given in Section \ref{sec:conclude}.

\section{Problem Formulation} \label{sec:problem}


This section describes the $q$-Composition scheme   
and proposes a corresponding integer programming formulation that can be solved using standard optimization software packages. In what follows, it is assumed that they underlying network is connected and that a given pair of sensor nodes are connected by an edge if they are within proximity of one another.

\subsection{$q$-Composite key distribution scheme}

\begin{figure}[h]
    \centering
    \includegraphics[width=0.60\textwidth]
    {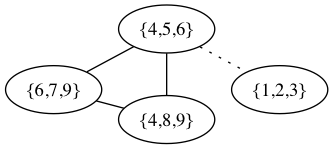}
    \caption{An illustration of Eschenauer and Gligor's scheme}
    \label{Fig:single_key}
\end{figure}

A special case of the $q$-Composite key distribution scheme for cases when $q=1$ is known as the Eschenauer and Gligor's method~\cite{Eschenauer_2002}, which is summarized a follows.
A key pool comprising a number of keys is randomly generated. Each key in the key pool is associated with a unique identifier. For each sensor node $i$, a subset 
of keys 
is selected from the key pool. The chosen subset 
is also called the {\em key ring} of node $i$. 
If two sensor nodes are within proximity and have at least one common key in their key rings, then they can communicate securely without relying on any intermediary nodes for exchanging information.  
The procedure  that utilizes the key identifiers for discovering common keys between two nodes is referred to as {\em shared key discovery} \cite{chen_survey_2014}.
Alternatively, if the key rings of two nodes do not share any common keys, they can (usually) communicate through a procedure called {\em key path establishment}. In this case, the nodes communicate indirectly through a sequence of secure links, where the sensors at the two endpoints of any given link share a common key. 
Figure \ref{Fig:single_key} illustrates the Eschenauer and Gligor's method on a simple network with a predetermined key rings. For instance, the key ring associated with node $A$ is given by $\{6,7,9\}$.  
Observe that sensor nodes $A$, $B$,
and $C$ are within proximity of one another, and the key rings corresponding to any given pair share a common key.  Hence, each of the three sensors can directly communicate with one another.  However, although sensor $D$ is within proximity of sensor $B$, they cannot communicate directly due to the fact that their key rings do not contain a common key. Consequently, sensor $D$ becomes an isolated node of the network.

\begin{figure}[h]
    \centering
    \includegraphics[width=0.60\textwidth]
    {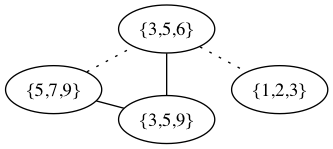}
    \caption{An illustration of $q$-Composite scheme with $q=2$}
    \label{Fig:double_key}
\end{figure}

As in the Eschenauer and Gligor's method, the $q$-Composite scheme uses the same procedure for distributing keys among the sensor nodes, for shared key discovery, and for establishing key paths. However, two nodes that are within proximity must have at least $q$ common keys in their key rings in order to communicate. In this sense, the $q$-Composite scheme offers a higher level of cryptographic security in that multiple keys may be required to establish inter-sensor communications.
Figure \ref{Fig:double_key} illustrates the mechanism 
of the $q$-Composite scheme when $q=2$. Notice that sensors $A$ and
$B$ only have one common key (i.e., key $5$), thus they are not able to communicate directly as their respective key rings must share at least 2 common keys. However, sensors $A$ and $B$ can communicate indirectly ``through'' sensor $C$ via key path establishment.

\subsection{A mathematical programming representation of the $q$-Composite key distribution scheme}

In this section, we  discuss the assumptions and objective of 
the defined $q$-Composite key distribution problem, and introduce 
a corresponding the integer programming model. 
Given a network whose topology is known, assume that each sensor node has 
a limited storage capacity that restricts the number of keys it can store.  
To assign keys in a highly secure manner, they should be distributed so 
as to avoid assigning the same $q$ common keys to more than one node 
within any given sensor's proximity. Additionally, the overuse of any given 
key is restricted by limiting the number of times it can be assigned to the nodes. 
Altogether, the emergent optimization problem involves maximizing the 
number of sensors within proximity of one another that can communicate 
(i.e., share at least $q$ keys), all while satisfying the sensors' memory 
capacities and the required communication security level. We next present 
an integer program with quadratic constraints that can provide optimal key 
management policies for the $q$-Composite key distribution scheme 
under the defined setting.

\subsubsection{An integer programming formulation} 

Let $G = (V, E)$ be an undirected, simple, connected graph that represents the 
underlying sensor network,  where each vertex is a sensor node, and an edge 
$(i,j) \in E$ indicates that vertices $i$ and $j$ are within proximity of one another. 
Let the set $N(i) = \{j : (i,j)\in E, \forall j \in V \}$ contain the vertices adjacent to 
vertex $i \in V$. Denote $K$ as the set of randomly generated keys that define 
the key pool, $m_k$ as the amount of memory required to store key $k \in K$, 
$c_i$ as the memory capacity limit of vertex $i \in V$, and $t_k$ as the maximum 
number of vertices that key $k$ can be assigned to. Finally, we define a parameter 
$p \in [0, 1]$ to adjust the security ``level'' by regulating the number of times a key 
can be used for communications by any given vertex.

For a given graph $G$ and a set of keys $K$, define the following decision variables:
\[
x_{ik} = \begin{cases}
1, & \text{if key $k \in K$ is assigned to node $i \in V$} \\
0, & \text{otherwise},
\end{cases}
\]
\[
z_{ij} = \begin{cases}
1, & \text{if edge $(i,j) \in E$ have $q$ common keys} \\
0, & \text{otherwise}.
\end{cases}
\]
Then, the described $q$-Composite KMP can be formulated as follows:
\begin{subequations}\label{equ:kd}
\begin{align} 
\max \limits_{x, z} \quad &  \sum \limits_{(i,j) \in E} z_{ij} \label{equ:obj} \\ 
\st \quad & \sum_{k \in K} m_k x_{ik}  \leq c_i, \ \ \ \forall \, i \in V  \label{equ:1cons} \\
\quad & \sum_{k \in K} x_{ik} x_{jk} \ge q z_{ij}, \ \ \ \forall \, (i,j) \in E  \label{equ:2cons} \\ 
\quad & \sum_{j \in N(i)} x_{ik}x_{jk} \le p |N(i)| + \alpha, \ \ \ \forall \, i \in V,  \forall k \in K  \label{equ:3cons} \\
\quad & \sum_{i \in V} x_{ik} \le t_k,  \ \ \ \forall \, k \in K \label{equ:4cons} \\  
\quad & x_{ik} \in \{0,1\}, \ \ \ \forall \, i \in V, \ \forall \, k \in K \label{equ:5cons} \\
\quad & z_{ij} \in \{0, 1\}, \ \ \ \forall \, (i,j) \in E, \label{equ:6cons} 
\end{align}
\end{subequations}
where the objective (\ref{equ:obj}) maximizes the number of vertices that can directly communicate, thereby maximizing the communication effectiveness of the network. Constraints (\ref{equ:1cons}) enforce that the key ring of any given vertex $i \in V$ does not exceed its memory capacity limit $c_i$.  Constraints (\ref{equ:2cons}) determine whether a given vertex pair $(i,j)$ that are within proximity share at least $q$ common keys. For a given vertex $i \in V$ and key $k \in K$, constraint (\ref{equ:3cons}) restricts the number of times that $k$ can be used by $i$ to encrypt communications with vertices in its neighborhood $N(i)$. The parameter $p \in [0,1]$ represents a percentage, while $\alpha$ is a positive integer that is adjusted to ensure that $p|N(i)|+\alpha$ is greater or equal to 1. Finally, to prevent excessive use of a given key $k \in K$, constraints (\ref{equ:4cons}) prevents $k$ from being assigned more than $t_k$ times among all the vertices.

The quadratic terms in constraints (\ref{equ:2cons}) and (\ref{equ:3cons})
can be linearized using standard techniques from binary programming. 
Consider a quadratic term $w_1 w_2$, where $w_1, w_2 \in \{0,1\}$. 
Then, $w_1 w_2$ can be equivalently represented by introducing a binary variable $y_{12} \in \{0,1\}$ and the following auxiliary inequalities:
\begin{subequations}
\begin{align} 
y_{12} \leq w_1, \ y_{12} \leq w_2 \\
y_{12} \ge w_1 + w_2 - 1.
\end{align} 
\end{subequations}

\noindent Analogously, by replacing each quadratic term $x_{ik}x_{jk}$ in Constraints (\ref{equ:2cons}) and (\ref{equ:3cons}) with a binary variable $y_{(i,j)}^k$, and introducing the necessary auxiliary inequalities, an equivalent integer linear program takes the form
\begin{subequations}\label{equ:kd-ref}
\begin{align} 
\max \limits_{x, y, z} \quad &  \sum \limits_{(i,j) \in E} z_{ij} \label{equ:obj2}\\ 
\st \quad & (\ref{equ:1cons}), \ (\ref{equ:4cons}) - (\ref{equ:6cons})  \nonumber \\
\quad & \sum_{k \in K} y_{(i,j)}^k \ge q z_{ij}, \ \ \ \forall \, (i,j) \in E  \label{equ:1cons2} \\ 
\quad & \sum_{j \in N(i)} y_{(i,j)}^k \le p |N(i)| + \alpha, \ \ \ \forall \, i \in V, \forall k \in K  \label{equ:2cons2} \\
\quad & y_{(i,j)}^k \leq x_{ik}, \ y_{(i,j)}^k \leq x_{jk}, \ \ \ \forall \, (i,j) \in E, \forall k \in K  \label{equ:3cons2}\\
\quad & y_{(i,j)}^k \geq x_{ik} + x_{jk}  - 1, \ \ \ \forall \, (i,j) \in E, \forall k \in K  \label{equ:4cons2} \\
\quad & y_{(i,j)}^k \in \{0,1\},  \ \ \ \forall \, (i,j) \in E, \forall k \in K.  \label{equ:5cons2}
\end{align}
\end{subequations}

Although the introduced model assumes particular ``security'' constraints and an objective function that govern how the keys are distributed, various extensions and modifications can easily be introduced.  Namely, the inequalities \eqref{equ:3cons} and \eqref{equ:4cons} can be modified (or removed) to meet the application at hand. For instance, rather than preventing a key $k \in K$ from being assigned to vertices within the immediate neighborhood of vertex $i \in V$, inequalities \eqref{equ:3cons} can be modified to prevent assignment of $k$ within a specified distance of $i$.
Furthermore, since the objective function in model \eqref{equ:kd} increases the likelihood that a key path can be established between any pair of vertices by maximizing the number of adjacent vertices that share at least $q$ keys, additional constraints requiring that all vertices within proximity of each other must share $q$ or more keys can be included. At the cost of additional complexity, this enhancement would guarantee the existence of a key path between any pair of vertices. Alternative variants of model \eqref{equ:kd} may involve designing key distribution strategies that maximize the number of vertices that would need to be compromised so as to obtain all the keys used for network-wide communications.  Finally, for settings where the network topology is not known in advance, model \eqref{equ:kd} can be extended to consider graphs with randomly changing topologies.  Depending on the requirements for secure key distributions and the availability of historical data, optimization techniques including robust optimization, stochastic optimization, and dynamic optimization can be applied to tackle the underlying stochastic settings. Such modeling extensions and corresponding solution algorithmic techniques will be pursued in subsequent studies.

\section{Case studies} \label{sec:exp}


Numerical experiments analyzing the effectiveness and solution performance of the proposed model were conducted.  Problem \eqref{equ:kd-ref} was solved for cases when $q = 1, 2$ for randomly generated Erdos-R\'enyi graphs of orders $|V| = 10, 30, 50, 100$ with average densities in the range $[0.04, 0.5]$. For any graph configuration of order $|V|$ and density $d$, one hundred instances were generated and solved. Depending on the particular graph instance, the memory capacity limit of vertices was set to $c_i = 5,6,7,8$, and the total number of times a given key $k \in K$ could be assigned was $t_k = 3,4,5$. The larger values of $c_i$ and $t_k$ were assigned when solving graph configurations of the larger orders. Throughout the numerical experiments, we let $m_k = 1, \forall k \in K$. Computational time limits were imposed based on the value of parameter $q$ (shown below). 

For each configuration, the average computational times of the instances that solved within the time limit were reported. Whenever an optimal solution was not obtained for all the instances of a configuration, the average computational time was reported for the instances that were solved within the time limit, while the average optimality gap was reported for the remaining instances. All computations were conducted on 64 core Intel(R) Xeon(R) CPU E7- 8837  @ 2.67GHz CentOS computer with 1TB of RAM and implemented in Python 2.7 and Gurobi 7.0.1. Each problem instance was solved using 1 core.

\subsection{Numerical experiments and results for the $1$-Composite key distribution scheme}

Thirteen graph configurations with various combinations of  $|V|$ in the range $[10,100]$ and $d$ in the range $[0.03, 0.5]$ were considered for $q=1$.  For each configuration, Table \ref{Tab:expset1} lists values for $|V|$, $d$ and parameters $|K|$, $q$, $p$, $c_i$, $t_k$; while Table \ref{Tab:params1} presents the number of instances that solved within the time limit, and the average solution times and/or average optimality gaps. The computational time limit for each instance was set to 7200 seconds.  The symbol ``--'' was used to report the solution time for configurations where the computational time limit was exceeded for all one hundred instances.

\begin{figure}[h!]
    \centering
    \includegraphics[width=0.85\textwidth]{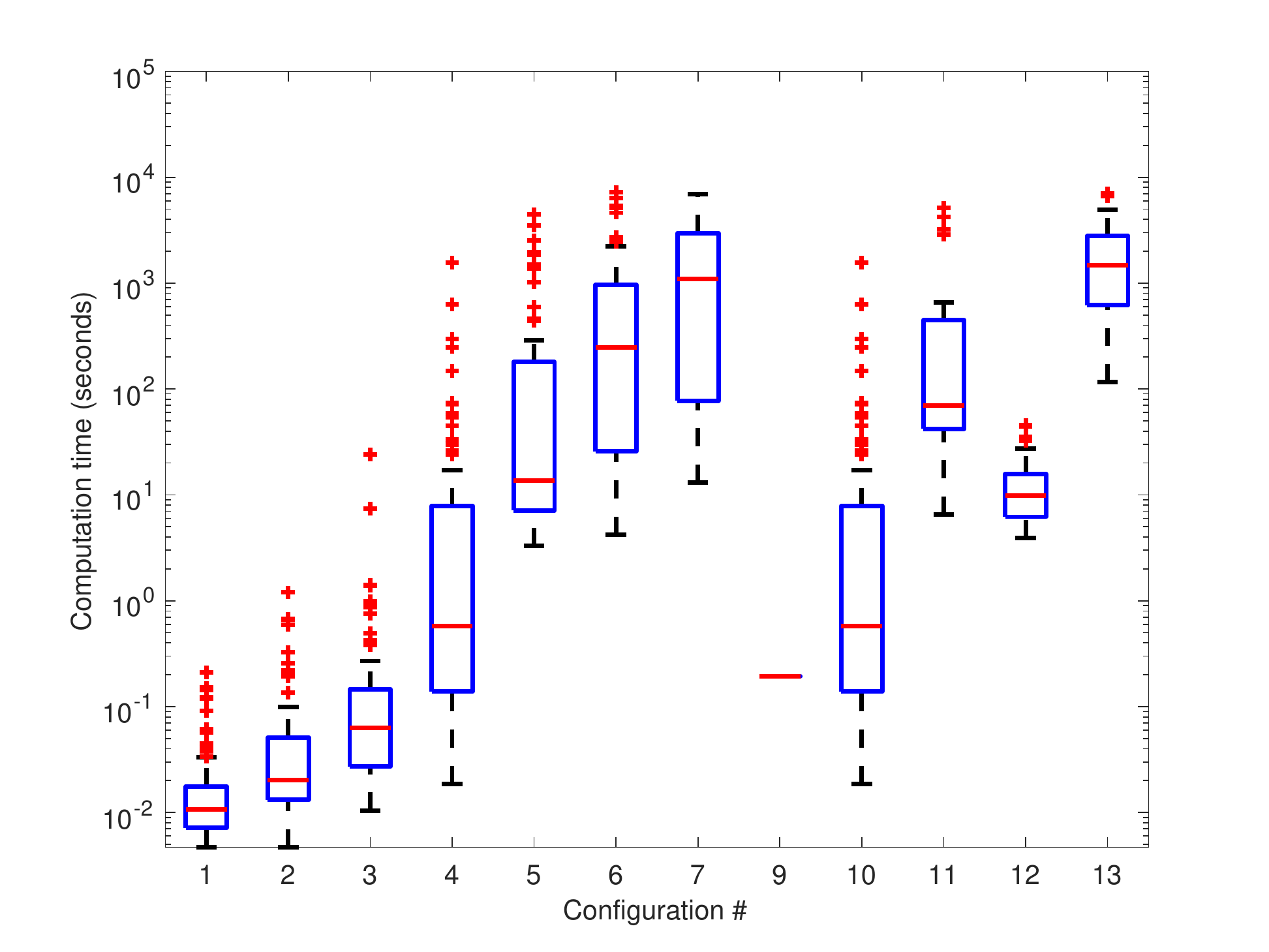}
    \caption{A boxplot of computation times (in seconds) for instances solved within the 7200 second time limit when $q=1$}
    \label{Fig:time_q1}
\end{figure}

\begin{figure}[h!]
    \centering
    \includegraphics[width=0.85\textwidth]{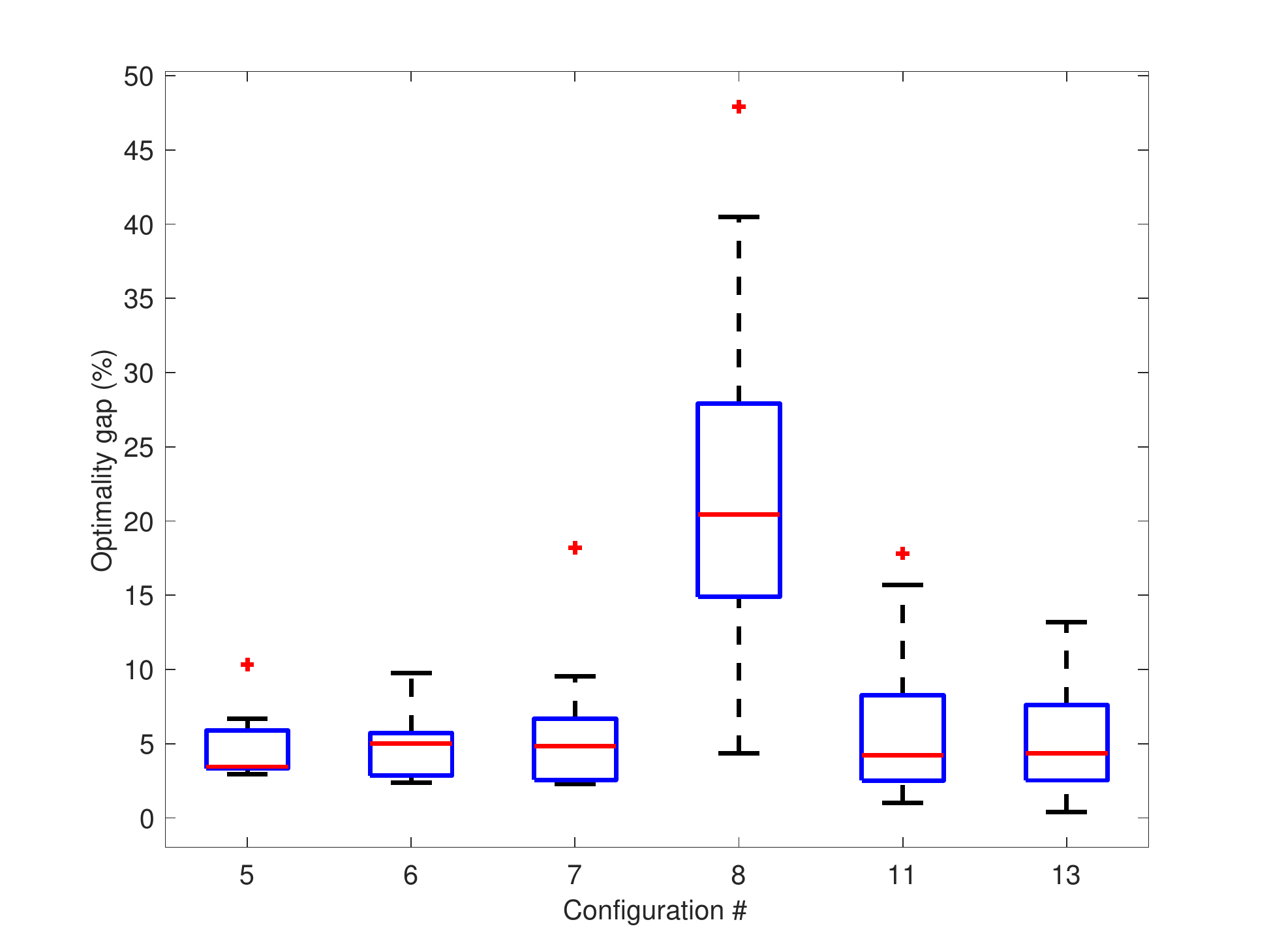}
    \caption{The boxplot for the optimality gap of the instances terminated because of reaching time limit for the example of $q=1$}

      \label{Fig:gap_q1}
\end{figure}

For a given order $|V|$, Table~\ref{Tab:params1} demonstrates that the solution time significantly increases as the graph density increases. For graph configuration instances where optimal solutions were found within the time limit, Figure~\ref{Fig:time_q1} furnishes a box plot depicting  statistics of computation times. It clearly demonstrates that the computation time increases with higher density values given that all other parameters remaining fixed (see Configurations 1-4 and Configurations 5-7). Furthermore, when the number of vertices increases, the computation time increases significantly. Due to the fact that Gurobi only solved two instances of Configuration 8 within the time limit, the corresponding statistics are not reported in Figure~\ref{Fig:time_q1}. Also, note that the box plot for Configuration 9 is relatively narrow. This resulted from the application of sparse graph instances, which, in turn, enabled Gurobi to obtain optimal key distributions very quickly. Clearly, the computational effort is heavily influenced by the density and the size of the underlying graph. Statistics for the optimality gaps are illustrated in Figure~\ref{Fig:gap_q1} for the instances where an optimal solution was not found within the time limit.  With the sole exception of Configuration 8, which comprised instances with the highest tested density, the results indicate that the optimality gaps were generally similar over all the configurations. Notably, the optimality gaps are relatively small for the tested configurations\footnote{
Note that Gurobi may produce a nonzero optimality gap even if the incumbent solution is optimal. In such cases, additional time is required to validate that the incumbent solution is optimal}.

\begin{table}[htbp]
\centering
\begin{tabular}{c|ccccccc}
Configuration \# & $|V|$ & $d$  & $|K|$ & $q$ & $p$ & $c_i$ & $t_k$ \\ \hline
1                & 10    & 0.2  & 10    & 1   & 0.3 & 5     & 3     \\
2                & 10    & 0.3  & 10    & 1   & 0.3 & 5     & 3     \\
3                & 10    & 0.4  & 10    & 1   & 0.3 & 5     & 3     \\
4                & 10    & 0.5  & 10    & 1   & 0.3 & 5     & 3     \\
5                & 30    & 0.05 & 20    & 1   & 0.3 & 6     & 3     \\
6                & 30    & 0.08 & 20    & 1   & 0.3 & 6     & 3     \\
7                & 30    & 0.10 & 20    & 1   & 0.3 & 6     & 3     \\
8                & 30    & 0.15 & 20    & 1   & 0.3 & 6     & 3     \\
9                & 50    & 0.04 & 30    & 1   & 0.4 & 7     & 4     \\
10               & 50    & 0.05 & 30    & 1   & 0.4 & 7     & 4     \\
11               & 50    & 0.08 & 30    & 1   & 0.4 & 7     & 4     \\
12               & 100   & 0.03 & 60    & 1   & 0.4 & 8     & 5     \\
13               & 100   & 0.05 & 60    & 1   & 0.4 & 8     & 5    
\end{tabular}

\caption{Parameter settings for the example of $q=1$}
\label{Tab:expset1}
\end{table}

\begin{table}[htbp]
\centering

\begin{tabular}{c|c|c|c}
Configuration \# & \# instance solved & Average time (s) & Average gap (\%) \\ \hline
1                & 100                & 0.02             & 0\%              \\
2                & 100                & 0.07             & 0\%              \\
3                & $100$              & 0.46             & 0\%              \\
4                & $100$              & 36.24            & 0\%              \\
5                & $62$               & 377.15           & 4.43\%           \\
6                & $51$               & 1026.71          & 4.48\%           \\
7                & $42$               & 1705.12          & 5.12\%           \\
8                & $2$                & 454.07           & 21.10\%          \\
9                & $100$              & 0.61             & 0\%              \\
10               & $100$              & 2.54             & 0\%              \\
11               & $20$               & 858.33           & 5.80\%           \\
12               & $100$              & 12.56            & 0\%              \\
13               & $57$               & 1994.77          & 4.99\%          
\end{tabular}
\caption{Computation statistics for the example of $q=1$}
\label{Tab:params1}
\end{table}

\subsection{Numerical experiments and results for the 2-Composite key distribution scheme}

\begin{figure}[h!]
    \centering
    \includegraphics[width=0.85\textwidth]{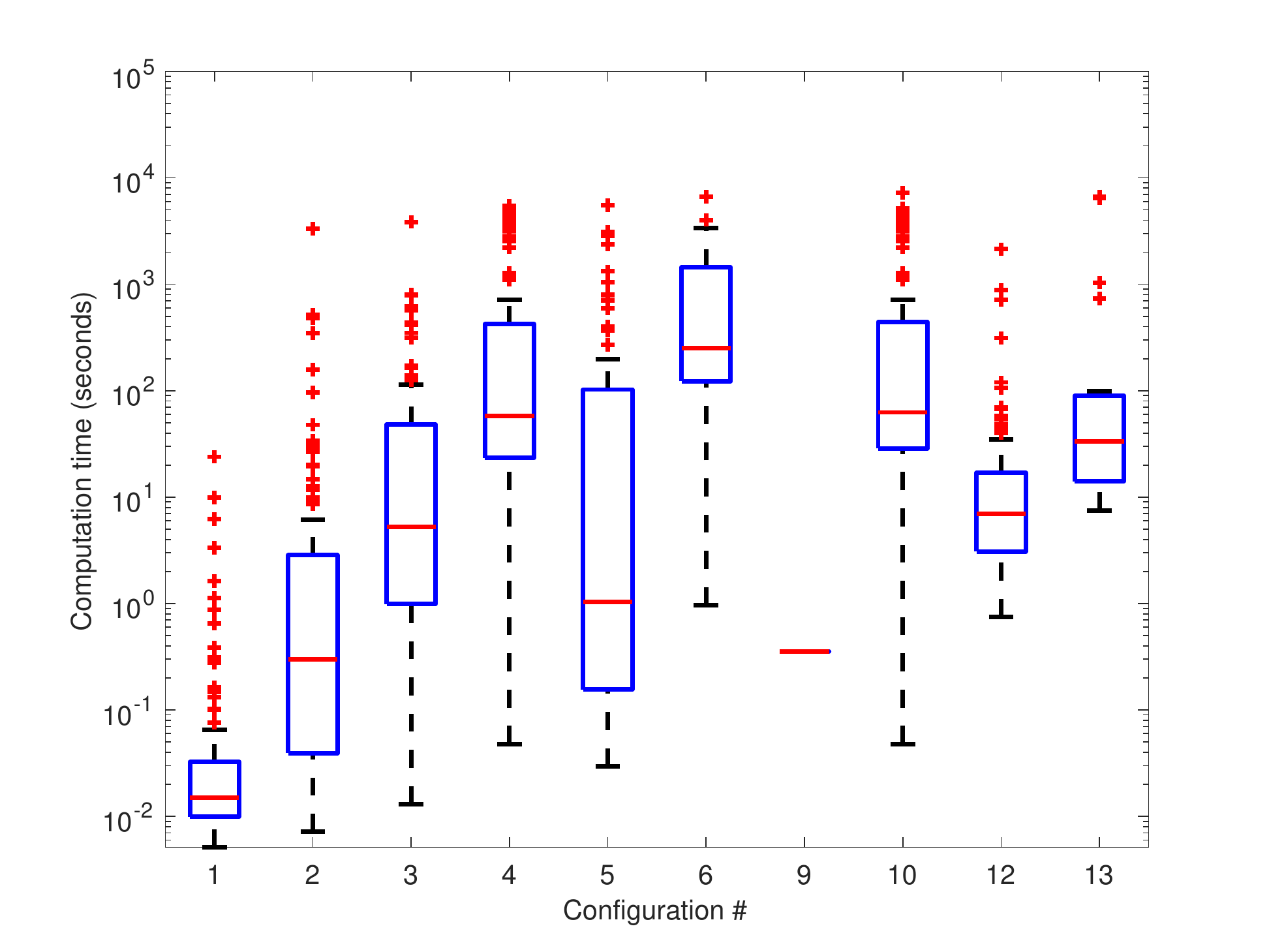}
    \caption{A boxplot of computation times (in seconds) for instances solved within the 10800 second time limit when $q=2$}
    \label{Fig:time_q2}
\end{figure}

\begin{figure}[h!]
    \centering
    \includegraphics[width=0.85\textwidth]{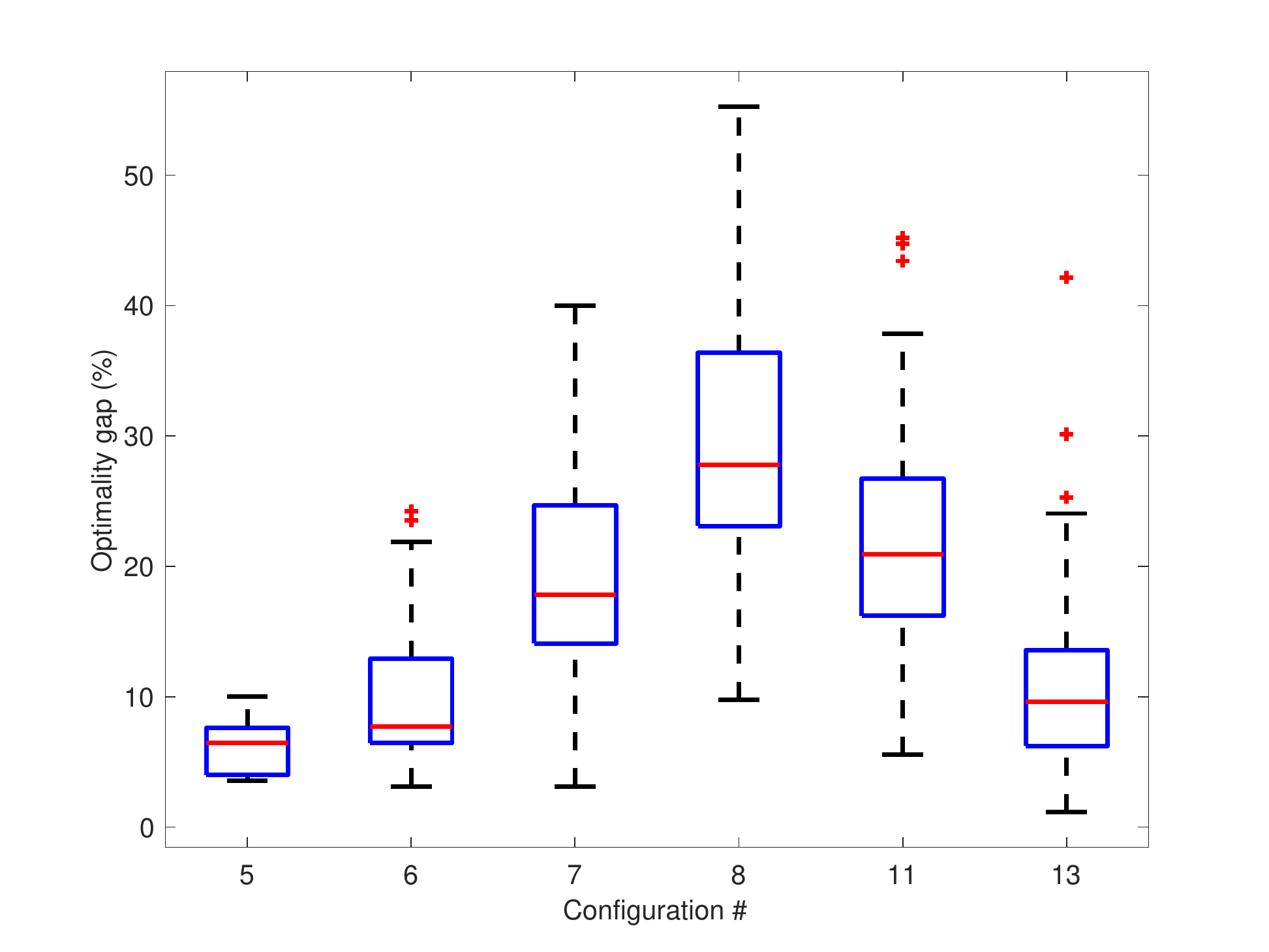}
    \caption{The boxplot for the optimality gap of the instances terminated because of reaching time limit for the example of $q=2$}
    \label{Fig:gap_q2}
\end{figure}

For the case when $q=2$, thirteen graph configurations with various combinations of  $|V|$ in the range $[10,30]$ and  $d$ in the range $[0.15, 0.5]$ where considered. The computational time limit for each instance was set to 10800 seconds. As previously, the graph configuration details and parameter values are shown in Table~\ref{Tab:expset2}; while the number of instances solved, average solution times, and average optimality gaps are reported in Table~\ref{Tab:params2}.  Figures~\ref{Fig:time_q2}-\ref{Fig:gap_q2} presents box plots illustrating summary statistics about the average solution times and average optimality gaps, respectively. 
Similar to the case of $q=1$, the computation times increase when the number of vertices or/and density increases. However, the case of $q=2$ is generally more computationally intensive.
For the computation times, similar trends can be observed as for $q=1$. However, the case of $q=2$ is generally more computationally intensive. This can be observed for Configurations 7, 8, and 11, where Gurobi could not solve any of the associated instances within the time limit.  



\begin{table}[htbp]
\centering
\begin{tabular}{c|ccccccc}
Configuration \# & $|V|$ & $d$  & $|K|$ & $q$ & $p$ & $c_i$ & $t_k$ \\ \hline
1                & 10    & 0.2  & 10    & 2   & 0.4 & 5     & 4     \\
2                & 10    & 0.3  & 10    & 2   & 0.4 & 5     & 4     \\
3                & 10    & 0.4  & 10    & 2   & 0.4 & 5     & 4     \\
4                & 10    & 0.5  & 10    & 2   & 0.4 & 5     & 4     \\
5                & 15    & 0.2  & 15    & 2   & 0.4 & 6     & 4     \\
6                & 15    & 0.3  & 15    & 2   & 0.4 & 6     & 4     \\
7                & 15    & 0.4  & 15    & 2   & 0.4 & 6     & 4     \\
8                & 15    & 0.5  & 15    & 2   & 0.4 & 6     & 4     \\
9                & 25    & 0.15 & 25    & 2   & 0.5 & 7     & 5     \\
10               & 25    & 0.2  & 25    & 2   & 0.5 & 7     & 5     \\
11               & 25    & 0.3  & 25    & 2   & 0.5 & 7     & 5     \\
12               & 30    & 0.15 & 30    & 2   & 0.5 & 8     & 5     \\
13               & 30    & 0.2  & 30    & 2   & 0.5 & 8     & 5    
\end{tabular}
\caption{Parameter settings for the example of $q=2$}
\label{Tab:expset2}
\end{table}

\begin{table}[htbp]
\centering
\begin{tabular}{c|c|c|c}
Configuration \# & \# instances solved & Average time (seconds) & Average gap (\%) \\ \hline
1                & $100$               & 0.51             & 0\%              \\
2                & $100$               & 52.82            & 0\%              \\
3                & $99$                & 108.29           & 5.56\%           \\
4                & $99$                & 737.41           & 12.00\%          \\
5                & $85$                & 260.91           & 6.02\%           \\
6                & $27$                & 995.58           & 9.70\%           \\
7                & $0$                 & $-$              & 18.60\%          \\
8                & $0$                 & $-$              & 29.74\%          \\
9                & $100$               & 1.53             & 0\%              \\
10               & $94$                & 116.19           & 4.40\%           \\
11               & $0$                 & $-$              & 21.89\%          \\
12               & $95$                & 56.8             & 4.54\%           \\
13               & $20$                & 773.88           & 10.51\%         
\end{tabular}
\caption{Computation statistics for the example of $q=2$}
\label{Tab:params2}
\end{table}

\subsection{An illustrative example of the $2$-Composite key distribution scheme}

\begin{figure}[h!]
    \centering
 \includegraphics[width=1.0\textwidth]{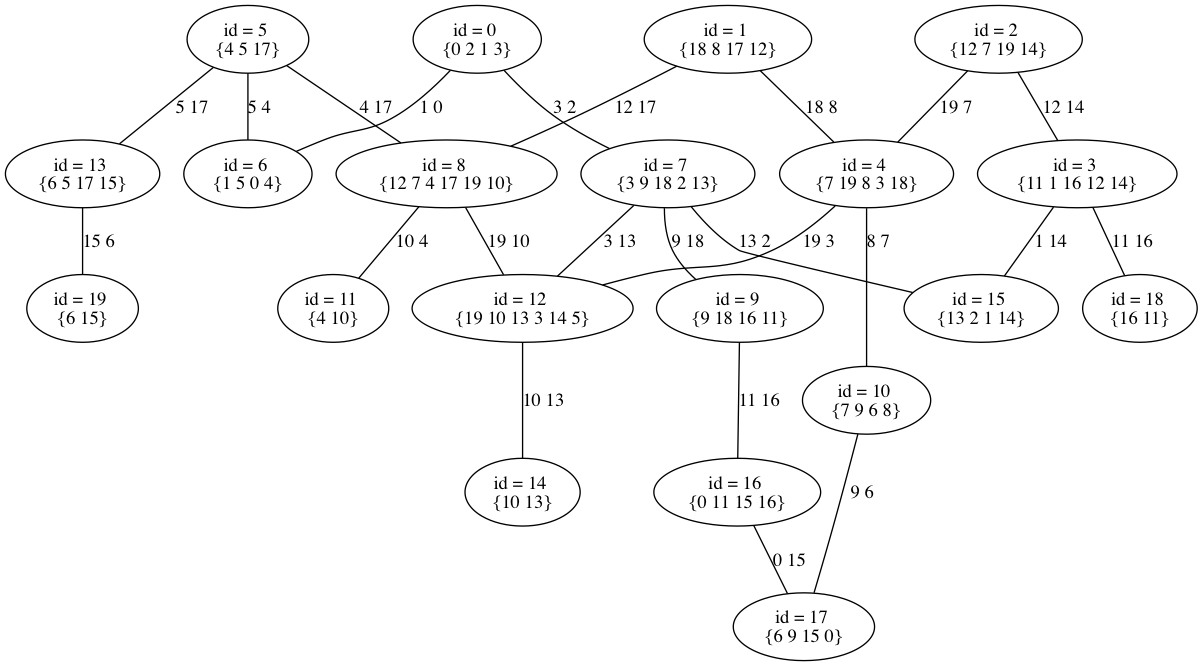}
    \caption{An optimal key distribution example obtained by solving model \eqref{equ:kd-ref} for the case of $q=2$ for a graph with $|V|=20$, $|E|=23$, and $|K|=20$}
     \label{Fig:example}
\end{figure}

Model \eqref{equ:kd-ref} was solved for the graph in displayed Figure \ref{Fig:example} consisting of $|V|=20$, $|E|=23$. By way of deducing meaningful experimental parameter values, we evoke the Advanced Encryption Standard (AES) for constrained wireless sensor nodes as described in \cite{Didla:2008:OAE:1390576.1390581}, where the authors focused on devices like the Crossbow Mica2 that have 4 KB RAM, and Texas Instruments MSP430 chips that have up to 10 KB RAM. 
According to AES, key lengths can be 128, 192, or 256 bits (16, 24, 32 bytes). 
Indeed, the implementation in \cite{Didla:2008:OAE:1390576.1390581} employed sensors with 260 bytes of RAM and 5160 bytes of ROM, where a vertex storing six AES keys in its RAM would require consumption of 192 bytes of memory.
To this end, we let the vertex memory capacity limits were $c_i = 6$, $\forall i \in V$. The key pool was fixed at $|K|=20$ and $t_k = 4$, $\forall k \in K$. 

Benefits of the developed mathematical programming-based protocol can be demonstrated by comparison against the naive fully pairwise distribution scheme, which would require two shared keys between each communicating vertex pair such that no key is used more than once. Observe that the solution produced by model \eqref{equ:kd-ref} enables direct or indirect (via key paths) communications between all vertex pairs.  However, assuming that the memory capacity limit of any given vertex does not exceed $m_k$, when implementing a naive pairwise distribution scheme, a minimum spanning tree (not shown) of the graph in Figure \ref{Fig:example} would reveal that the key pool size would require eighteen additional keys in order to establish communications via the corresponding nineteen edges.



\section{Concluding Remarks} \label{sec:conclude}

In this paper, we propose an integer programming approach for the key management problem in wireless sensor networks with emphasis on the {\em $q$-Composite key distribution} scheme for assigning secret keys to sensor nodes. Numerical experiments demonstrating the effectiveness and computational efforts required to solve the proposed model are conducted.  Experimental results indicate that the developed method works effectively for small sized networks.

To enable the identification of optimal key management policies in larger networks, subsequent studies will focus on developing effective solution methods for the presented model. Additionally, the authors' will consider other variants of key distribution protocols under the scope of mathematical programming, as well as stochastic extensions that address setting when the topology of the underlying wireless sensor network is not known in advance.

{
	\paragraph{Acknowledgements.}  
	The work presented in this paper was carried out while Guanglin Xu 
	was a postdoctoral fellow at the Institute for Mathematics and its Applications during the IMA's annual 
	program on {\em Modeling, Stochastic Control, Optimization, and Related Applications}.
}

\bibliographystyle{plain}

\bibliography{kd}

\begin{thebibliography}{10}

\bibitem{Blom1985}
Rolf Blom.
\newblock {\em An Optimal Class of Symmetric Key Generation Systems}, pages
  335--338.
\newblock Springer Berlin Heidelberg, Berlin, Heidelberg, 1985.

\bibitem{Blundo1993}
Carlo Blundo, Alfredo De~Santis, Amir Herzberg, Shay Kutten, Ugo Vaccaro, and
  Moti Yung.
\newblock {\em Perfectly-Secure Key Distribution for Dynamic Conferences},
  pages 471--486.
\newblock Springer Berlin Heidelberg, Berlin, Heidelberg, 1993.

\bibitem{Camtepe_2007}
S.~A. Camtepe and B.~Yener.
\newblock Combinatorial design of key distribution mechanisms for wireless
  sensor networks.
\newblock {\em IEEE/ACM Transactions on Networking}, 15(2):346--358, April
  2007.

\bibitem{chen_survey_2014}
Chi-Yuan Chen and Han-Chieh Chao.
\newblock A survey of key distribution in wireless sensor networks.
\newblock {\em Security Comm. Networks}, 7(12):2495--2508, December 2014.

\bibitem{Didla:2008:OAE:1390576.1390581}
Shammi Didla, Aaron Ault, and Saurabh Bagchi.
\newblock Optimizing aes for embedded devices and wireless sensor networks.
\newblock In {\em Proceedings of the 4th International Conference on Testbeds
  and Research Infrastructures for the Development of Networks \& Communities},
  TridentCom '08, pages 4:1--4:10, ICST, Brussels, Belgium, Belgium, 2008. ICST
  (Institute for Computer Sciences, Social-Informatics and Telecommunications
  Engineering).

\bibitem{Eschenauer_2002}
Laurent Eschenauer and Virgil~D. Gligor.
\newblock A key-management scheme for distributed sensor networks.
\newblock In {\em Proceedings of the 9th ACM Conference on Computer and
  Communications Security}, CCS '02, pages 41--47, New York, NY, USA, 2002.
  ACM.

\bibitem{he_achieving_2006}
Tian He, Pascal Vicaire, Ting Yan, Qing Cao, Gang Zhou, Lin Gu, Liqian Luo,
  Radu Stoleru, John~A. Stankovi, and Tarek Abdelzaher.
\newblock Achieving {Real}-{Time} {Target} {Tracking} {Using} {Wireless}
  {Sensor} {Networks}.
\newblock Technical report, 2006.

\bibitem{Kim_2007}
Sukun Kim, Shamim Pakzad, David Culler, James Demmel, Gregory Fenves, Steven
  Glaser, and Martin Turon.
\newblock {\em Health monitoring of civil infrastructures using wireless sensor
  networks}, pages 254--263.
\newblock 2007.

\bibitem{Liu_2003}
Donggang Liu and Peng Ning.
\newblock Establishing pairwise keys in distributed sensor networks.
\newblock In {\em Proceedings of the 10th ACM Conference on Computer and
  Communications Security}, CCS '03, pages 52--61, New York, NY, USA, 2003.
  ACM.

\bibitem{ruj_pairwise_2013}
S.~Ruj, A.~Nayak, and I.~Stojmenovic.
\newblock Pairwise and {Triple} {Key} {Distribution} in {Wireless} {Sensor}
  {Networks} with {Applications}.
\newblock {\em IEEE Transactions on Computers}, 62(11):2224--2237, November
  2013.

\bibitem{Simplicio_2010}
Marcos~A. Simpl\'{\i}cio, Jr., Paulo S. L.~M. Barreto, Cintia~B. Margi, and
  Tereza C. M.~B. Carvalho.
\newblock A survey on key management mechanisms for distributed wireless sensor
  networks.
\newblock {\em Comput. Netw.}, 54(15):2591--2612, October 2010.

\bibitem{Fundamentals.2010}
Dargie Waltenegus and Christian Poellabauer.
\newblock {\em Wiley: {Fundamentals} of {Wireless} {Sensor} {Networks}:
  {Theory} and {Practice} - {Waltenegus} {Dargie}, {Christian} {Poellabauer}}.
\newblock Wiley, 2010.

\bibitem{werner_allen_1607983}
G.~Werner-Allen, K.~Lorincz, M.~Ruiz, O.~Marcillo, J.~Johnson, J.~Lees, and
  M.~Welsh.
\newblock Deploying a wireless sensor network on an active volcano.
\newblock {\em IEEE Internet Computing}, 10(2):18--25, March 2006.

\end{thebibliography}

\end{singlespace}
\end{document}